\documentclass[preprint2]{aastex}

\shorttitle{Life-bearing Planet Search}
\shortauthors{S. V. W. Beckwith}

\begin{document}

%% LaTeX will automatically break titles if they run longer than one line. However, you may use \\ to force a line break if you desire.
\rm
\title{Detecting Life-bearing Extra-solar Planets with Space Telescopes \\ {\small Revised version}}


\author{Steven V. W. Beckwith\altaffilmark{1,2,3}}
\affil{\it University of California, 1111 Franklin St., Oakland, CA 94607-5200, USA}
\email{\tt steven.beckwith@ucop.edu}

%% affiliation1

\altaffiltext{1}{Space Telescope Science Institute}
\altaffiltext{2}{Johns Hopkins University}
\altaffiltext{3}{University of California}


% abstract
\begin{abstract}
One of the promising methods to search for life on extra-solar planets (exoplanets) is to detect its signature in the chemical disequilibrium of exoplanet atmospheres. Spectra at the modest resolutions needed to search for methane, oxygen, carbon dioxide, or water will demand large collecting areas and large diameters to capture and isolate the light from planets in the habitable zones around the stars. Single telescopes with coronagraphs to isolate the light from the planet will have to be 8\,m or more in diameter to generate sample sizes with a reasonable probability of finding at least one life-bearing planet; interferometers of smaller telescopes can overcome some of the limitations but will still need large similarly large collecting areas. Even larger telescopes will be needed to detect atmospheric signatures in transiting planets. In all cases, the sample sizes increase as the third power of telescope diameter. Direct observation using coronagraphs or interferometers are most sensitive to planets around stars with masses similar to that of the Sun, whereas transit observations favor low-mass stars near the nuclear burning limit. If the technical difficulties of constructing very large space telescopes can be overcome, they will be able to observe planets near hundreds to thousands of stars with adequate resolution and sensitivity to look for the signatures of life. 
 \end{abstract}
 

%------------ body of article ------------------->>

\section{INTRODUCTION}
The detection of more than 200 planets outside the Solar System is a powerful incentive to search for extra-terrestrial life. Although extra-terrestrial life could take on many guises, economy of hypothesis (and a practical approach) implies that we should first search for signs of life similar to those seen on Earth. An advantage of using the Earth as a proxy for analysis is that it bounds the problem and provides concrete examples of signatures subject to passive detection, i.e. not requiring signals broadcast by sentient beings. 

In the Earth's present atmosphere, the chemical components have been altered by life. Evidence for life on Earth could be detected from afar in the spectral signatures of these molecules: oxygen, carbon dioxide, methane, and water vapor \citep{sag93}. The rise of photoplankton and plants on Earth created an atmosphere with a large reservoir of oxygen today that requires steady production by photosynthesis to maintain its present level. If we could study the same spectral signatures in the atmospheres of exoplanets, we could search for signs of life similar to some of the earliest and most robust forms on our own planet. But the Earth's atmosphere has been altered by life several times over the last 4\,Gyr, and there are potentially many signatures that could indicate the presence of life on exoplanets \cite{tra02, sea02, kalt07}. Chemical signatures of life on other planets would revolutionize our thinking about Earth's uniqueness and provide tantalizing evidence that we are not alone in the universe.

Observing exoplanets directly is difficult owing to their proximity to the much brighter stars that keep them warm. Although technically challenging, this problem is well understood, and there are a variety of strategies that can reduce the brightness of the starlight without diminishing the light from the planet for direct detection \citep{guy06,cas06} or use the star itself as a background source to probe the atmosphere when the exoplanet transits the face of the star \citep{cha02, ehr06}. The first technique must overcome diffraction in the pupil of the telescope, a well understood phenomenon, and it is possible to characterize the detection problem in general terms to understand the kinds of instruments that will be needed to study exoplanets and search for signs of life. The second technique depends only on the photometric accuracy of an observation and is easy to calculate for any star.

Any planet supporting life as on Earth must satisfy two broad criteria: (1) it must have surface temperatures in the range 273 to 373\,K, where water is in the liquid phase, and (2) it must have an atmosphere. The first criterion is met if the planet is in the {\it habitable zone} (HZ) around the star, a range of orbital distances where the equilibrium temperature for a rotating body is between the freezing and boiling points of water. The second criterion is met if the planet is rocky and can retain an atmosphere; current estimates specify a mass between 0.5 to 10 Earth masses. Smaller planets will not retain their atmospheres, and larger planets accrete gas and become gas giants. These criteria are necessary but not sufficient to create Earth-like life. Although they are probably far too restrictive to encompass all the possibilities for other life forms in the universe---or even on Earth itself---they are the only ones amenable to remote observation with technology that we can foresee at present and thus provide a good basis for a targeted search.

There have been many calculations aimed at refining our ideas of the habitable zone \citep[e.g.][]{kast93}, suitable samples of stars to search \citep{tur03, tur04, tur06}, and the impact of specific telescopes on such a search \citep{alg07}, and there is some disagreement about the likelihood of success depending on the different assumptions used. Most authors concentrate on photometric detection alone, ignoring the means to search for life (e.g. Agol 2007). 

But the utility of photometric searches alone to identify exoplanets for subsequent study could be obviated by the difficulty of measuring the orbits accurately enough to recover them at a later time \cite{bro05, bro07}. For apertures sizes under discussion for the Terrestrial Planet Finder (TPF) mission, habitable zones of most potential target stars are blocked by the coronagraph. The planets of interest will spend the majority of time behind the central obscuration of the imaging instrument. To ensure efficient recovery of a newly discovered planet at future observing epochs, it will be necessary to estimate the orbit to high accuracy from a small number of astrometric measurements. This requirement implies a lower limit to the aperture size based on operational requirements (Brown et al. 2007). Achieving adequate astrometric precision may demand apertures larger than any so far discussed for TPF (Brown 2007, personal communication). If true, discovery and immediate spectroscopy of candidate sources near the stars may be the most efficient means of identifying those most interesting for follow up observations to determine if life is, indeed, present, making it essential to understand the spectroscopic requirements at the outset. 

The purpose of this article is to derive the main scaling parameters for the study of life-bearing exoplanets in known samples of stars to understand the size of the telescopes needed for a robust search. We adopt simple but optimistic assumptions to bound the problem and find the minimum size for survey telescopes. Using only the lowest order approximations and assuming ``best case'' observing conditions allows robust conclusions about the scale of facilities needed to tackle the search for life. 

The main premise is that direct photometric detection of exoplanets in a band where the exo-planet atmosphere is free of chemical signatures cannot be the endpoint of any mission to search for life-bearing planets; spectra of the atmospheres will be the major advance of observing the planets directly. Moreover, the rapid increase in the number of exoplanets discovered to date suggests that finding terrestrial planets will be easiest with indirect methods, such as observing radial velocity, photometric or astrometric variations in the host stars, and the real thrust of direct observations will be to  search for signs of life. 

\section{Detecting Earth}
\label{sec:detectingearth}
The Earth's flux density viewed from a distance is reasonably well constrained by its size and the requirement that its equilibrium temperature is determined by external illumination at the temperature of the Sun. The same statements apply to exoplanets in the habitable zones around distant stars. An important result is that the flux densities of exoplanets in the habitable zones across a broad spectrum are roughly independent of the stellar luminosity and temperature. 

If the exoplanet is a spherical body with radius, $r_{p}$, and temperature, $T_{p}$, and it is viewed at a distance, $D \gg r_{p}$, it's thermal emission is at most that of a blackbody, $B$:
\begin{eqnarray}
F_{pt}(\nu) & = & {\pi r_{p}^{2} \over D^{2}} B(\nu, T_{p}),
\label{eqn:exothermal}
\end{eqnarray}
The total emission is $4 \pi r_{p}^{2} \sigma T_{p}^{4}$. 

\begin{figure}[htb]
\plotone{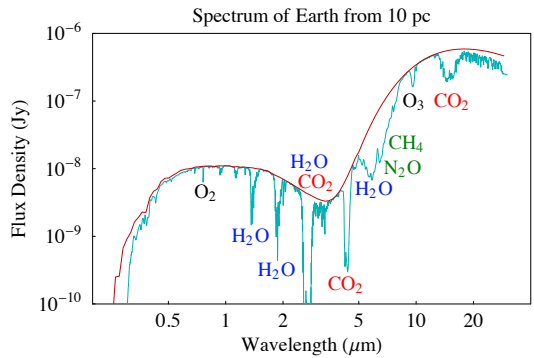}
\caption{The lower line is a synthetic spectrum of Earth viewed from 10\,pc to illustrate the spectral features in the Earth's present-day atmosphere. The upper line is a combination of the solar irradiance and a Planck function at 286\,K.}
\label{fig:earthspectrum}
\end{figure}

The reflected light is in principle a rather complicated function of the planet's orbital phase and the detailed scattering properties of its atmosphere \cite[e.g.][]{sea00a}. For the purposes of this paper, it will be sufficient to assume the planet scatters like a Lambert sphere viewed at quadrature, the viewing angle $\alpha = 90^{\circ}$, with a fraction $1- a(\nu)$ of the light absorbed in the process. If it is illuminated by a star of radius, $r_{*}$, with an effective temperature, $T_{*}$, at orbital distance, $R_{p}$, from the star, the observed flux density of the reflected light is:
\begin{eqnarray}
F_{pr}(\nu) & = & \frac{2}{3} {r_{p}^{2}\over R_{p}^{2}} \phi(\alpha) a(\nu) { \pi r_{*}^{2}\over D^{2}} \pi B(\nu, T_{*}), \ \ \ \  \\
& = & \frac{2}{3} a(\nu) {r_{p}^{2}\over R_{p}^{2}} { r_{*}^{2}\over D^{2}} \pi B(\nu, T_{*}), 
\label{eqn:exoreflect}
\end{eqnarray}
where the Lambert sphere has a geometrical cross section of $\frac{2}{3}$ and a phase function $\phi(\alpha) = \frac{1}{\pi} (\sin\alpha + (\pi - \alpha) \cos\alpha) = \frac{1}{\pi}$ for $\alpha = \frac{\pi}{2}$ \cite{rus16, sea00a}. The unabsorbed fraction, $a(\nu)$, will in general depend on frequency. For the Earth, it varies between $\sim 0.7$ near 0.3\,$\mu$m to less than $0.3$ longward of about 0.6\,$\mu$m and depends on variable factors such as cloud cover and exposed land mass \cite{kast93, tur06, mcc06}. We adopt $a(\nu) = a =0.5$ for the rest of this paper to produce a bright planet that should be most easily observed. As shown below, uncertainties in $a(\nu)$ will not affect the estimates in this paper significantly. The actual spectra would also include radiative transfer effects in the atmosphere and show absorption lines from molecules (see Fig.~\ref{fig:earthspectrum}).

%Figure 1

Combining (\ref{eqn:exothermal}) and (\ref{eqn:exoreflect}) with the assumptions above, the flux density of the planet is:
\begin{eqnarray}
F_{p}(\nu) & \approx & {\pi r_{p}^{2}\over D^{2}}\left( {1\over 3} {r _{*}^{2}\over {\rm R_{p}}^{2}} B(\nu, T_{*}) + B(\nu, T_{p})\right). \ \ \ \ \ 
\label{eqn:fnuplanet}
\end{eqnarray}
This relation neglects some interesting details, such as the variable atmospheric transmission and radiative transfer effects as well as adopting assumptions about the albedo and phase function that are at best educated guesses. It turns out that these simplifications do not affect the overall energy balance enough to change the conclusions of this article, but they are, nevertheless, crucial for understanding the chemical composition of the planet's atmosphere.

At 10\,pc, the apparent magnitude of the Earth at visual and near infrared wavelengths from (\ref{eqn:fnuplanet}) is about 29(AB), $\sim10$\,nJy, too faint for spectra with any existing telescopes. 

The principal sources of noise in detecting exoplanets will be local zodiacal light, any equivalent exo-zodiacal light in the other planetary system, any residual light from the star not cancelled by the instrument, and the planet itself. Since we are interested in the limits of nature, we assume that starlight is completely eliminated by the telescope and instrument and consider only the other contributions.  

\begin{figure}[bht]
\plotone{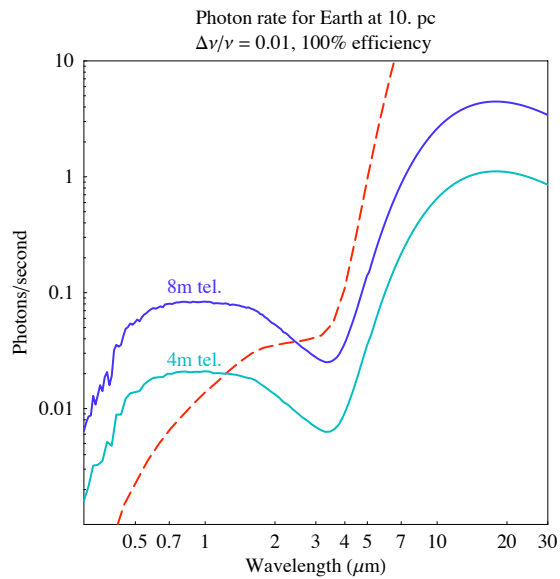}
\caption{The photon rates of an exo-Earth viewed from 10\,pc distance with 4\,m  and 8\,m telescopes, together with the photon rates for the local zodiacal light for a diffraction-limited telescope.}
\label{fig:exophotonrates4and8}
\end{figure}

The zodiacal light has the same spectrum as the Sun at wavelengths shorter than about 3\,$\mu$m, but reduced by the optical depth of the zodiacal cloud in the direction of observation, and it is approximately a blackbody function at thermal infrared wavelengths with a temperature that depends somewhat on the viewing angle with respect to the Sun. The important point is that both the zodiacal light and exo-zodiacal light light will be uniform surface brightnesses to a good approximation whose observable fluxes depend only on the area--solid angle product (\'etendue) of the telescope and not the distance to the source. The spectrum of the zodiacal light combines a visual optical depth, $\tau_{\rm z} \sim 10^{-7}$, and thermal emissivity, $\epsilon_{z}(\nu) \sim 0.5$, both strong functions of ecliptic line of sight, to obtain the brightness of the zodiacal light as:
\begin{eqnarray}
I_{\rm z}(\nu) & \approx & \tau_{z} [ B(\nu, T_{\odot}) +  \epsilon_{z} B(\nu, T_{\rm zody}) ]. \ \ \ \ \ 
\label{eqn:zody}
\end{eqnarray}
It is straightforward to calculate the exact zodiacal emission in any direction from models in the literature (e.g. Leinert et al. 1998), and as a practical matter we will adopt a viewing angle at solar elongation $90^{\circ}$ and ecliptic latitude $45^{\circ}$ as ``typical'' of a large survey seeking to minimize zodiacal background: $I_{\rm zody}(\nu)$ is about 0.3\,MJy\,sr$^{-1}$ at 0.5\,$\mu$m and 10\,MJy\,sr$^{-1}$ at 10\,$\mu$m. Figure~\ref{fig:exophotonrates4and8} compares the photon rates seen by 4 and 8\,m telescopes from an Earth 10\,pc away with the zodiacal light for an assumed resolving power of 100.

It is impossible to know how strong the contribution from the exo-zodiacal light is. It is easy to find stars with very large quantities of interplanetary dust \cite{bei06} and stars without detectable dust emission, so this term is one of the larger uncertainties in our estimates as stressed by Agol (2007). The best case is if there is no exo-zodiacal light at all. Alternatively, if the other planetary system is identical to the Solar System, the amount of exo-zodiacal light is just twice the local value. The easy way to understand this factor of 2 is to realize that we view the exo-zodiacal cloud through its entire thickness, whereas we reside at the mid-plane of the local zodiacal light (within 3$^{\circ}$) and thus look out through only half the optical depth in any direction. The exo-zodiacal cloud will also be well resolved by any telescope that can block the starlight and still observe the planet. For the moment, we assume the contribution of the exo-zodiacal light is negligible.

%Figure 2

The noise of an observation will be dominated by the (Poisson) statistics of the incoherent photon stream from the planet and the zodiacal background. Denoting photon rates as $\dot{N}_{\gamma}(p)$ and $\dot{N}_{\gamma}(z)$, respectively, the signal to noise ratio, $SN$, after observing time, $t$, is:
\begin{eqnarray}
SN & = & {\eta \dot{N}_{\gamma}(p) t \over \left(  \eta \dot{N}_{\gamma}(p) t +  \eta \dot{N}_{\gamma}(z) t \right)^{1/2}}, \\ 
      & = & \sqrt{ \eta t {1 \over h } { \Delta \nu \over \nu} }  { A_{\rm tel} F_{p}(\nu)  \over \sqrt{ A_{\rm tel} \left( F_{p}(\nu)+ \Omega_{\rm tel} I_{z}(\nu)\right) } }, \nonumber \\
\label{eqn:SNcalc1}
\end{eqnarray}
with the area and solid angle of the telescope denoted $A_{\rm tel}$ and $\Omega_{\rm tel}$, and the overall detection efficiency $\eta$. For a circular telescope that is diffraction-limited, the \'etendue, $A_{\rm tel} \Omega_{\rm tel}$, is just equal to the square of the observing wavelength,  $\lambda^{2}$. In the limit where the planet is far away from the observer (us), the photon noise is dominated by the zodiacal light regardless of whether it is local or around the exo-planet, since its contribution is independent of both source distance and telescope diameter, but the photon flux from the exo-planet is proportional to $D^{-2}$. Equation (\ref{eqn:SNcalc1}) may then be written as:
\begin{eqnarray}
 SN & = &  t^{1/2} \sqrt{ {1 \over h } {\Delta \nu \over \nu}  \eta} { {\pi \over 4} d_{\rm tel}^{2} F_{p}(\nu)  \over \sqrt{ \left({c \over \nu}\right)^{2} I_{z}(\nu)} }, \label{eqn:SNcalc2} \\
  & \propto & t^{1/2} \eta^{1/2}\, \left({d_{\rm tel} \over D}\right)^{2}.  \label{eqn:SNcalc3} 
 \end{eqnarray}
The expression in (\ref{eqn:SNcalc3}) shows the explicit dependence of $SN$ and observation time on telescope diameter and source distance in the faint limit. If exo-zodiacal light is comparable in brightness to the local value, $SN$ diminished by a factor of order $\sqrt{3}$. It is useful to note that even though the amount of exo-zodiacal light is uncertain, it introduces only a modest uncertainty into the signal-to-noise calculation considering many of the other uncertainties inherent in the planet detection problem. However, we will assume it is zero in subsequent calculations to consider the best case scenarios for taking spectra of planets.

\begin{figure}[htb]
\plotone{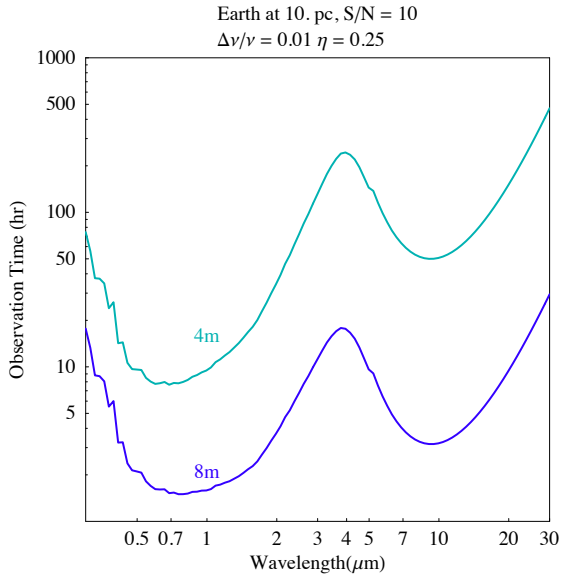}
\caption{Integration time to achieve S/N=10 for each spectral element for a practical 4\,m and 8\,m telescope at a resolving power of 100 assuming only background from the local zodiacal light; integration times would increase for any exo-zodiacal emission. The overall detection efficiency, $\eta$, is 0.25 for this figure.}
\label{fig:exointegrationtime3}
\end{figure}

Recasting (\ref{eqn:SNcalc2}) in terms of observation time for Earth at a distance is especially useful for the planet detection problem. Using (\ref{eqn:fnuplanet}) and (\ref{eqn:zody}),  $F_{p}$ and $I_{z}$ can be written in terms of assumptions about the star, planet, and observing system parameters. To simplify later calculations, we define $SN_{10} \equiv SN / 10$, $\Gamma_{100} \equiv {\nu / \Delta\nu} / 100$,  $d_{8} \equiv d_{\rm tel} / 8\,$m, $D_{10} \equiv D / 10\,$pc, and $t_{24} \equiv t / 24\,$hr:
\begin{eqnarray}
 t_{24}  & \approx & SN_{10}^{2} \eta^{-1}\, \Gamma_{100} \left({D_{10} \over 4 d_{8}}\right)^{4} \ \ \ \   
\label{eqn:timecalcnum}
\end{eqnarray}
assuming an observation wavelength $\lambda = 1\,\mu$m, justified below.  Equation~(\ref{eqn:timecalcnum}) shows that the time for an observation increases as the fourth power of distance over telescope diameter when observing a point source against a uniform brightness background in the photon limit. Figure~\ref{fig:exointegrationtime3} plots the times need to take low signal-to-noise spectra of an Earth at 10\,pc with 4 and 8\,m telescopes. This strong dependence means that observations of Earth-like exoplanets with a telescope of a given size will be effectively limited to some maximum distance: 
\begin{eqnarray}
D_{s} & = & 40\,{\rm pc}\  d_{8} \left({\eta t_{24}\over \Gamma_{100} SN_{10}^{2}}\right)^{1/4} \\
& \approx & 30\,{\rm pc}\  d_{8} 
\label{eqn:distsense}
\end{eqnarray}
where we adopted $\eta = 0.25$ and set the other parameters to unity for the purposes of later simplification. This equation applies to observations of reflected light at $1\,\mu$m from the planet and is strictly valid only for distances where the planet is fainter than the zodiacal light. For bright objects, the source flux itself also adds to the noise further increasing the required integration time. The limiting distance for longer wavelengths using (\ref{eqn:SNcalc2}) will have the same functional dependence but a smaller distance normalization given by the relative increase in the integration time at the longer wavelength (Fig.~\ref{fig:exointegrationtime3}) to the negative one-fourth power.

One of the interesting results of this calculation is that the observation time near the peak of the thermal radiation, $\sim 10\,\mu$m, is about the same as at shorter wavelengths despite the enormously larger background radiation. The reason is because the background noise increases only as the square root of the photon rate, and very high photon rates reduce the relative noise from the background very rapidly, at which point the high photon rate from the exoplanet is easy to detect. The only real disadvantage of thermal infrared observations is the poor diffraction-limited resolution of a single filled aperture, making it difficult to detect planets at small angular distances from the stars. 

The choice of 1\,$\mu$m for (\ref{eqn:timecalcnum}) is appropriate for any observations with a single-telescope, where the ability to eliminate starlight is tied to the telescope diameter as discussed below. Observations in the thermal infrared at wavelengths longer than a few microns encounter enormous challenges of starlight suppression at small angular distances from the star, making single telescope observations unworkable. However, multiple telescopes used as an interferometer or the combination of a single telescope with a distant occulter would separate the observational sensitivity considered here from the suppression of starlight. We will consider both cases in the following sections to show the independent effects of sensitivity and starlight suppression.

%Figure 3

The assumptions used thus far assume a perfectly diffraction-limited telescope where the foreground radiation is dominated by zodiacal light. These conditions are easy to meet for a space telescope; they are extremely difficult to meet for any telescope on the Earth's surface. Even if the distortion of the wavefront from the Earth's atmosphere could be entirely compensated by an adaptive optics system, the atmosphere radiates strongly and absorbs light from the exoplanet in all the lines of interest in Fig.~\ref{fig:earthspectrum} making the characterization problem almost impossible for a ground-based telescope.

The practical effect of (\ref{eqn:timecalcnum}) is to limit the useful distance for observations of an exo-Earth with a single telescope. It takes about  24\,hr to take a spectrum of Earth at a distance of 40\,pc with an 8\,m telescope, a distance which shrinks to 27\,pc, if the detection efficiency is 25\%, typical for a good spectrograph. For a 4\,m telescope, the limiting distances for a 24\,hr observation time are 20 and 13.5\,pc, respectively, since the distance scales linearly with telescope diameter for this problem. These sizes are larger than the current generation of space telescopes. It is also useful to note the size of telescope below which the zodiacal light begins to dominate the light from the planet as a function of the distance to the planetary system, $F_{p}(\nu) \le \Omega_{\rm tel} I_{zody}(\nu)$ in the denominator of (\ref{eqn:SNcalc1}). Using (\ref{eqn:fnuplanet}) and (\ref{eqn:zody}) at a wavelength of 1\,$\mu$m, $d_{\rm tel} \le 3.25\,{\rm m}\ (D / 10\,{\rm pc})$. At thermal infrared wavelengths, zodiacal light dominates that light from any interesting exoplanet for all telescopes less than a few tens of meters in diameter.

An interferometer can be treated as a telescope with a sparsely-filled aperture. The total collecting area is just the sum of the individual collecting areas. Each telescope collects background radiation over a solid angle set by the diffraction-limit of each individual aperture, the ``main beam,'' modified by the interference pattern created by the spatial separation of the telescopes. 

One way to think about this problem is to note that the solid angle is proportional to the Fourier transform of the pupil-plane (u-v) coverage, which must be equivalent to the pupil-plane coverage of a single telescope with an area equal to the sum of the areas of the elements. Therefore, if the sum of the collecting areas of the interferometer elements is $A_{tot}$, the sensitivity of a single observation can be calculated using (\ref{eqn:SNcalc2}) with $d_{\rm tel} = \sqrt{4 A_{tot} / \pi}$. If a complete aperture synthesis is required to take spectra with the equivalent resolution of a single telescope with diameter, $d_{1}$, the total time will be $\pi d_{1}^{2} / (4 A_{tot})$ times the single observation time. Full synthesis may not be needed if the position of the exoplanet is known from some other observation. 

Sample volumes of the local neighborhood increase as the cube of the distance. Because there is a premium on maximizing the number of survey stars that can host potentially habitable planets, these results show a strong preference for large telescopes or interferometers with large collecting areas. The preferred size will depend on the density of suitable stars in the Solar neighborhood and their luminosity distribution to set the scale for the size of the habitable zones, the subject of the next section.

\section{Sample Sizes for Direct Observations}
\label{sec:directsamples}
The interesting candidate stars will be those whose habitable zones can be studied after the light from the star is suppressed to a level below the background. Because the starlight must be essentially eliminated---the Sun is $\sim 10^{10}$ times brighter than its reflected light from the Earth---starlight cancellation is incredibly demanding. It is generally a far higher technical hurdle than detection for exo-planet observations, and it is the reason that articles on this subject often underestimate the importance of background-limited sensitivity when considering exo-planet studies.

For a single telescope, starlight suppression requires a coronagraph that rejects light over a small area in the focal plane, while allowing observation of the light from outside this area. The smallest area that can be darkened depends on the limiting resolution of the telescope which can never be smaller than the diffraction limit of the pupil. An external occulter suppresses starlight by using a distant screen decoupled from the telescope to create a shadow of the starlight outside of which exoplanets may be studied. An interferometer is essentially a sparse aperture telescope that can be phased to null starlight like a coronagraph over an angular area the depends (inversely) on the separation of the elements. 

The {\it inner working angle}, $\theta_{IWA}$, is defined as the smallest angular separation between a bright star and a much fainter planet at which the planet can be studied with good cancellation of the starlight. For a single telescope with a circular pupil, the diffraction pattern is an Airy function \cite{BW99}, and $\theta_{IWA}$ cannot be less than the distance to the first null at $1.2\lambda / d_{\rm tel}$. Most realistic coronagraphs achieve inner working angles more than twice this angle, even with nearly optimal designs \cite{guy06}. The angle of the first null is typically many tens of milli-seconds of arc: $1.2\lambda / d_{\rm tel} =  0.21\,{\rm arcsec}\ \left({\lambda / 1\,\mu{\rm m}}\right) \left({d_{\rm tel} / 1\,{\rm m}}\right)^{-1}$.

For a telescope-occulter combination or an interferometer, the inner working angle is not constrained by the diameter of the light collecting telescope(s). An occulter produces a fixed inner working angle determined by its size and distance from the telescope. The inner working angle of an interferometer depends on the maximum distance between individual elements (telescopes) and will vary with position angle on the sky, depending its configuration. The resolution of an interferometer varies with wavelength in the same manner as a coronagraph, allowing us to characterize the inner working angle as a fraction of the diffraction-limit of the effective diameter of the telescope, $\theta_{IWA} = f_\theta 1.2 \lambda / d_{\rm tel}$ for any system, with $f_\theta \ge 1$ for a coronagraph, $f_\theta \ll 1$ for an interferometer, and $f_\theta(\lambda)$ a function of wavelength for an occulter but generally less than one for the wavelengths of observation. 

In all cases, the size of the star sample available for study will increase as the inner working angle decreases, making the smaller habitable zones around numerous low-luminosity stars observable at any distance. Put differently, as $\theta_{IWA}$ shrinks, the minimum mass of main sequence stars available for study decreases, and the sample size grows from the inclusion of low mass stars. 

We can simplify this problem by deriving scaling relationships for the size of the habitable zones around stars as a function of stellar parameters and characterizing the samples in terms of the present day mass function of stars that gives their number density as a function of stellar mass. The relative advantages of the different star suppression techniques will then emerge as a comparison of sample sizes for specific observing techniques.

Almost all the interesting spectral features in the Earth's atmosphere are longward of 0.7\,$\mu$m (Fig.~\ref{fig:earthspectrum}.) A spectrum should extend to at least $1\,\mu$m to be analyzed for evidence of disequilibrium chemistry indicative of life, and longer wavelengths bring in even more interesting features.  We adopt $\lambda = 1\,\mu$m as the minimum working wavelength to look for signs of life in spectra and note the desirability of observations to look for signatures of ozone, methane, and water that have strong bands throughout the thermal infrared beyond 10\,$\mu$m, where integration times are near a second minimum as seen in Fig.~\ref{fig:exointegrationtime3}.

The size of the habitable zone depends on the stellar luminosity. The total thermal emission, $4\pi r_{p}^{2} \sigma T_{p}^{4}$, must equal the total amount of absorbed starlight, $(1-a)L_{*}/(4\pi R_{p}^{2})$. The planet's temperature, $T_{p}$, is bounded between the freezing and boiling points of water, $273\,{\rm K} \le T_{p} \le 373\,{\rm K}$, a necessary but not sufficient condition for life as it is on Earth. We assume for this calculation that all the starlight is absorbed ($a=0$) to maximize the habitable zone radii. This condition is sufficient to compute the inner and outer habitable zone radii for a star of luminosity, $L_{*}$ using general formulae for the orbital radius, $R_{p}$, of a rotating planet at equilibrium temperature, $T_{p}$:
\begin{eqnarray}
R_{p} & = & \sqrt{L_{*} \over 16 \pi \sigma T_{p}^{4} } \label{eqn:orbitvslum}, \\
 & = & {1\over 2} \left({T_{*} \over T_{p}}\right)^{2} r_{*}. \label{eqn:orbitvstemp}
\end{eqnarray}
The inner and outer radii of the habitable zone are obtained from (\ref{eqn:orbitvstemp}) with $T_{p} = 373$ and 273\,K, respectively. Effects such as greenhouse warming or low absorption of starlight via high albedo will change these radii by modest factors \citep{kast93} in opposite directions; for example, the effective temperature of the Earth is $\sim255$\,K \cite{ort00} and the albedo is $\sim 0.5$ but the greenhouse effect keeps the surface warm. These corrections are not large enough to change the overall conclusions of this paper, however, and we will subsequently ignore them in the interest of simplicity. 

Using (\ref{eqn:orbitvslum}) to compute the radii bounding the habitable zone, $\theta_{IWA}$ determines the maximum angular area that can be blocked to achieve starlight suppression for a star at a known distance and luminosity. By examining large samples of stars, it is possible to determine how many have habitable zones that can be observed with a telescope or telescope system (interferometer or telescope and occulter).

To the accuracy required for the treatment in this paper, we express the stellar parameters in terms of the normalized stellar mass, $m \equiv m_{*} / M_{\odot}$, as follows:
\begin{eqnarray}
L_{*} & \approx & L_{\odot}\ m^{4} \label{eqn:luminosityvsmass} \\
r_{*} & \approx & r_{\odot}\ m \label{eqn:radiusvsmass} \\
T_{*} & \approx & T_{\odot}\ m^{0.5} \label{eqn:temperaturevsmass} 
\end{eqnarray}
The approximations are good to about 25\% for the local sample; for the Zero Age Main Sequence, the coefficients in front of the three equations would be $1.2 L_{\odot}$, $0.96 r_{\odot}$, and $1.1 T_{\odot}$, respectively, and the exponents would be 3.8, 0.92, and 0.49, respectively.

These relationships allow us to recast the equations for observing the atmospheres of exoplanets in terms of stellar mass. In particular, we can separate the impact on sample size of the collecting area from inner working angle, allowing any observing system to be optimized when trading off these two parameters.

The star is relevant only to the size of the habitable zone through (\ref{eqn:orbitvstemp}), the planet always having approximately the same spectrum (we will adopt $T_{\oplus} = 285$\,K to correspond to earlier sections):
\begin{eqnarray}
R_{p} & = & {1\over 2} \left({T_{\odot} \over T_{p}}\right)^{2} m^{2}\ r_{\odot}  \\
 & = & 1.0\,{\rm AU}\ \left({ T_{p} \over T_{\oplus}}\right)^{-2} m^{2}
\label{eqn:HZvsmass}
\end{eqnarray}
The size of the habitable zone is proportional to $\sqrt{L_{*}} \propto m^{2}$ from (\ref{eqn:orbitvslum}) and (\ref{eqn:luminosityvsmass}). For any inner working angle, there is a minimum mass star for which the habitable zone might be observed at any distance, $R_{p}(m_{\rm min}) = \theta_{IWA} D$.  We will specify the inner working angle as a multiple of the diffraction limit of a circular pupil, $\theta_{IWA} = f_{\theta} (1.2 \lambda_{\mu} / d_{\rm tel})$ and for generality define the equivalent diameter of the telescope system in terms of the collecting area, $d_{\rm tel} = \sqrt{4 A_{\rm tel} / \pi}$. Then, the maximum distance to a star of normalized mass, $m$, where the habitable zone can still be resolved is: 
\begin{eqnarray}
D_{IWA}(m) & = &  R_{p} / \theta_{IWA}   \\
 & \approx & 30\,{\rm pc}\, d_{8} f_\theta^{-1} \left({T_{p} \over T_{\oplus}}\right)^{-2} m^{2} \ \ \ \ \ 
\label{eqn:distIWA} 
\end{eqnarray}
For any star, the maximum distance for spectroscopic observations will be the lesser of $D_{IWA}$ or $D_{s}$, the limit from the integration time in (\ref{eqn:distsense}). The minimum stellar mass whose habitable zone is observable at distance, $D$, is:
\begin{eqnarray}
m_{\rm min} & = & \left({f_\theta \over d_{8}} {D \over 30\,{\rm pc}}\right)^{1/2} {T_{p} \over T_{\oplus}}
\label{eqn:mmin}
\end{eqnarray}

The present day mass function, $\xi(m)$, of local stars above the nuclear burning limit is well approximated by two power laws\footnote{Other choices are possible \cite{kro01}, but this choice fits both the 8\,pc sample and the more extended sample of solar mass stars to 25\,pc.} in stellar mass \cite{rei02}:
\begin{eqnarray}
\xi(m) & = & 0.02\,{\rm pc}^{-3}\  m^{-\alpha_{m}} 
\label{eqn:PDMF}
\end{eqnarray}
with $\alpha({0.08 \le m < 1.0}) = 1.1\approx 1$ and $\alpha({m \ge 1.0})$ $= 5.2 \approx 5$. The PDMF is defined such that the number density of stars in the mass range $m$ to $m+dm$ is $\xi(m)dm$. The large exponent for the upper end of the mass range describes the actual number density of stars above a few solar masses as opposed to the initial mass function which is derived by correcting for stellar evolution.

The sample size distribution for direct observations follows from (\ref{eqn:PDMF}) using the maximum volume for each stellar mass:
\begin{eqnarray}
d N_{d} / d m & = & {4 \pi \over 3} D^{3} \xi(m) 
\label{eqn:dnddm}
\end{eqnarray}
where $D$ is either $D_{IWA}$ or $D_{s}$. The stellar mass for which $D_{s}$ becomes the limiting distance depends on the rejection factor, $f_\theta$, and the place in the habitable zone where the planet resides, governed by $T_{p}$. If we assume a perfect coronagraph observing Earth, then $f_\theta (T_{p} / T_{\oplus}) \approx 1$, and $D_{s} < D_{IWA}$ for $m > 1$, exactly where the exponent in $\xi(m)$ changes value. The total sample for all masses above 0.1\,$M_{\odot}$ is:
\begin{eqnarray}
N_{d} & = & {4 \pi \over 3} \Big( \int_{0.1}^{1} D_{IWA}^{3}(m) \xi(m) dm \nonumber \\
& & \ \ \ \ \ \ \ \ +  D_{s}^{3}\int_{1}^{\infty} \xi(m) dm \Big) \label{eqn:Ntotal} \\
& \approx & 1900\, f_\theta^{-3} \left({T_{p} \over T_{\oplus}}\right)^{-6} d_{8}^{3} 
\label{eqn:Ntotal1}
\end{eqnarray}

The sample size depends strongly on the effective diameter of the telescope, inner working angle, and working wavelength in the low-mass limit. It depends weakly on the overall detection efficiency, integration time, signal-to-noise ratio, and spectral resolving power.

\subsection{Samples for Coronagraphs}
\label{subsec:coronagraphs}
Direct observations of planets with coronagraphs tend to be limited by the inner working angle and benefit from the largest possible habitable zone. Evaluating (\ref{eqn:Ntotal1}) with $f_\theta=2$, the sample size as a function of telescope diameter is:
\begin{eqnarray}
N_{d} & \approx &  280\, \left({T_{p} \over T_{\oplus}}\right)^{-6} d_{8}^{3}
\label{eqn:Ncorona3}
\end{eqnarray}
An 8\,m telescope with an excellent coronagraph should have almost than 300 stars around which direct characterizations of Earth-like planets could be made spectroscopically, assuming every star where suitable for observation (e.g. no binaries). But the dependence on the third power of telescope diameter shows that the best one could do with a 4\,m telescope is about 35 stars, and this is the total sample without correction for multiple star systems and stars that might not be suitable for life. On the other hand, a very large telescope, 16\,m, say, could study thousands of stars to characterize terrestrial planets. 

\subsection{Samples for Compound Telescopes}
\label{subsec:compoundsample}
The samples can grow rapidly when $f_\theta$ is less than 1. Use of an external occulter should produce $f_\theta \sim 1$, and an interferometer could give $f_\theta \ll 1$, although the instantaneous coverage of the ({\it u, v}) plane is sparse and would be of limited value unless the precise location of the planet were well known at the time of the observation.

Evaluating (\ref{eqn:Ntotal}) for the case $f_\theta = 0.1$ and $T_{p} = T_{\oplus}$ yields: 
\begin{eqnarray}
N_{d} & \approx & 3300\, d_{8}^{3}, \ \ \ {\rm for}\ f_\theta = 0.1\  \label{eqn:Ninterferometer}
\end{eqnarray}
This is the sample a good interferometer might access. An occulter can reduce $f_\theta$ to less than 1 (e.g. Cash 2006), but the sample size grows rather slowly with decreasing $f_\theta$ owing to the sensitivity limit. The typical size of an occulter to produce $f_\theta \le 1$ coupled with its distance from the telescope to produce a small inner working angle will make a practical occulter difficult to use. 

These examples show that the gain from using a larger telescope comes about mainly from the rapid increase in sensitivity with aperture size, $d_{\rm tel}^{-2}$, and the rapid increase in sample size owing to the larger volume probed: $N \sim D^{3} \sim d_{\rm tel}^{3}$. The use of an external occulter helps mitigate the technical demands of creating a nearly perfect coronagraph, although it will drive some very strong technical demands of its own---alignment of a $\sim60$\,m structure at 400,000\,km from a space telescope and moving it around to sample different stars will be no mean feat. Large telescopes will still have an enormous advantage for searching for signs of life on exoplanets.  

\section{Spectra of Transiting Exoplanets}
\label{sec:transits}
It is possible to distinguish very slight differences in the light curve at different wavelengths as an exoplanet transits the face of its central star. These differences result from transmission spectrum of the star through the planet's atmosphere making the apparent size of the planet larger at wavelengths where the atmosphere is opaque. Charbonneau et al. (2002) and Vidal-Madjur et al. (2003, 2004) demonstrated the utility of this method by detecting sodium, hydrogen, oxygen, and carbon in the extended atmosphere of the transiting exoplanet, HD~209458b, using the STIS spectrograph on 2.4\,m Hubble Space Telescope. Tinetti et al. (2007) used the Spitzer Space Telescope in a similar way to detect H$_{2}$O in HD~189733b. Transits provide an alternative way to look for the signatures of life in exoplanets \cite{sea00}.

As a planet passes across the face of a star, the starlight dims by approximately the ratio of the area of the planet to that of the star, $\Delta_{p} \approx \pi r_{p}^{2} / \pi r_{*}^{2}$, where we ignore the modest effect of limb darkening. If the scale height of the atmosphere is $h_{atm}$ and an atmospheric absorption feature has the effect of blocking all light to $x$ scale heights above the surface, there will be a small difference in the depth of the light curve viewed at the wavelength of the feature:
\begin{eqnarray}
\Delta_{atm} - \Delta_{p} & = & {(r_{p} + x h_{atm})^{2} - r_{p}^{2} \over r_{*}^{2} } \label{eqn:depth1}, \\
  & \approx & {2 x h_{atm} r_{p} \over r_{*}^{2} }, \label{eqn:depth2}
\end{eqnarray}
where the approximation assumes $x h_{atm} \ll r_{p}$. For the Earth crossing the face of the Sun seen from afar, this an excellent approximation: $h_{atm} \approx 9$\,km,  $r_{p} \approx 6400$\,km, and $\Delta_{atm} -\Delta_{p} = 2 \times 10^{-7}$ for $x=1$. The scale height of the atmosphere depends on the density and temperature of the planet through the usual thermodynamic relation:
\begin{eqnarray}
h_{atm} & = & {k T_{p} \over \bar{\mu} g }, \label{eqn:scaleh1} \\
 & = & {3 k T_{p} \over 4 \pi G \bar{\mu} \rho r_{p} }, \label{eqn:scaleh2}
\end{eqnarray}
where $g$ is the gravitational acceleration at the planet's surface, $\bar{\mu}$ is the mean molecular mass, and $\rho$ is the mean density of the planet\footnote{The density of a planet increases slightly with mass as the interior equation of state changes \cite{sea07}, but the effect is small enough to be ignored at this level of approximation.}. Recasting (\ref{eqn:depth2}) in terms of the density:
\begin{eqnarray}
\Delta_{atm} - \Delta_{p} & \approx & x {3 k T_{p} \over 2 \pi G \bar{\mu} \rho r_{*}^{2} }. \label{eqn:depth3}
\end{eqnarray}
This result shows somewhat non-intuitively that the atmospheric signature for a planet in the habitable zone is independent of the planet's size; to first order, it depends only on the density and temperature, the latter of which lies between the freezing and boiling points of water, thus bounding the range of possibilities. The stronger gravitational field of larger planets decreases the scale height to compensate for the increase in the planet's radius.

The number of scale heights for the effective cross section, $x$, depends on the strength of the absorption lines. In the Earth's atmosphere, the strength of the absorption lines is normally expressed in terms of the vertical optical depth from the ground to space. For a molecule whose number density, $n_{i}$,  as a function of height, $z$, is exponential with scale height, $h_{atm}$, and absorption cross section per molecule, $\sigma_{i}$, the vertical optical depth is $\tau_{0} = \int_{0}^{\infty} n_{i} \sigma_{i} \exp (-z / h_{atm})\ dz = n_{i} \sigma_{i} h_{atm}$. Then the optical depth through the atmosphere along a line with an impact parameter $x$ scale heights above the surface is given by:
\begin{eqnarray}
\tau(x) & = &  \tau_{0} \int_{-\infty}^{\infty} e^{-\left(\sqrt{\left( x + r_{p}^{\prime} \right)^{2} + y^{2}} - r_{p}^{\prime}\right)}\ d y \ \ \ \ \ \  \\
 & = & 2 \tau_{0} e^{r_{p}^{\prime}} \left(r_{p}^{\prime}+x\right) {\rm K_{1}}(r_{p}^{\prime}+x), \\
 & \approx & \tau_{0} \sqrt{2\pi} \left(r_{p}^{\prime}+x\right)^{\frac{1}{2}} e^{-x}
\label{eqn:tauheight}
\end{eqnarray}
where $r_{p}^{\prime} = r_{p} / h_{atm}$, ${\rm K_{1}}$ is a modified Bessel function, and the approximation assumes $r_{p}^{\prime}+x \gg 1$. The value of $x$ that gives the effective atmospheric cross section for any absorption line seen in transit is the solution to (\ref{eqn:tauheight}) that gives $\tau(x) = 1$. For $r_{p}^{\prime} \gg x$, the solution is:
\begin{eqnarray}
x & \approx & 0.92 + {1\over2} \ln\left({r_{p} \over h_{atm}}\right) + \ln(\tau_{0}) \label{eqn:planetx} \\
 & \approx &  4.20 + \ln(\tau_{0})\ \ {\mbox for }\ r_{p}^{\prime} = \frac{6400}{9}.  
\label{eqn:earthx}
\end{eqnarray}

The lines shortward of 1\,$\mu$m in Fig.~\ref{fig:earthspectrum} have vertical optical depths, $\tau_{0}$, between 0.1 and 1 \cite{cri00}, meaning $x$ is between 2 and 4. There are very strong H$_{2}$O lines at wavelengths of a few microns, in principle giving $x \sim 10$, although for the Earth itself, water vapor is confined to the tropopause within a single scale height, so the lack of vertical mixing means that $x({\rm H_{2}O}) \sim 1$ despite the large $\tau_{0}$. The important result of (\ref{eqn:planetx}) is that the effective cross section within an absorption line depends on the logarithm of the line optical depth; even order of magnitude uncertainties in the total atmospheric density affect the transit depths modestly.

Detecting the absorption lines in a transit observation means achieving very high photometric accuracy in the difference spectrum in and out of the line. The best photometric accuracy that can be achieved during a transit is if all the starlight is detected with perfect efficiency and the only noise source is the statistical fluctuation of the photon rate as assumed in (\ref{eqn:SNcalc1}). Using the formalism above for measurement time, $t_{obs}$, and assuming the difference of two measurements:
\begin{eqnarray}
{\sigma_{N_{\gamma}} \over N_{\gamma}} & = &  \left( {2 \over \dot{N}_{\gamma} t_{obs} }\right)^{1/2}, \label{eqn:photom1} \\
 & = & \left( {\pi \over 8 h} {\Delta\nu \over \nu} F_{*}(\nu) d_{tel}^{2} t_{obs} \right)^{-1/2}. \label{eqn:photom2}
 \end{eqnarray}
The total observation time will be the time of a single transit times the number of transits. Viewed from a great distance, the maximum transit time is just $t_{trans} = 2 r_{*} / v_{p} = 2 r_{*} \sqrt{ R_{p} / G M_{*} }$, and the time for an average transit (i.e. not across the equator) is just one half this value. The number of transits observed during the lifetime of a space mission, $t_{m}$, is just $n_{P} = t_{m} / P $, with $n_{P}$ truncated to an integer . The total observation time is:
\begin{eqnarray}
t_{obs} & = & n_{P} t_{trans}  \\
 & = & t_{m} {1\over 2\pi} \sqrt{ G M_{*} \over R_{p}^{3}} 2 r_{*} \sqrt{R_{p} \over G M_{*}} \\
 & = & t_{m} {1\over \pi} {r_{*} \over R_{p}} \label{eqn:obstime} \\
  & = & t_{m} {2\over \pi} \left({T_{p} \over T_{*}}\right)^{2} \label{eqn:obstimetemp}
\end{eqnarray}
We used (\ref{eqn:orbitvstemp}) to express $R_{p}$ in terms of the temperatures to yield another non-intuitive result: the total observation time {\it decreases} as the stellar effective temperature {\it increases}. This effect combined with the lower contrast for a transit is more than enough to offset the advantage of having a brighter star with a higher photon rate:
\begin{eqnarray}
SN_{tr} & = & { \Delta_{atm} - \Delta_{p} \over {\sigma_{N_{\gamma}} / N_{\gamma}} } \label{eqn:sntrans0} \\
 & = &  x {3 k T_{p}^{2} d_{tel}  \over 4 \pi G \bar{\mu} \rho r_{*}^{2} T_{*}}  \left( {1 \over h} {\Delta\nu \over \nu} F_{*}(\nu) t_{m} \right)^{1/2}  \ \ \ \label{eqn:sntrans1} \\
& = &
x {3 k T_{p}^{2} d_{tel}  \over 4 \pi G \bar{\mu} \rho r_{*}^{2} T_{*}}  \left( {1 \over h \Gamma} t_{m} \right)^{1/2}  \nonumber \\
&& \times\ \left({\pi r_{*}^{2} \over D^{2}} {2 h \nu^{3} \over c^{2}} {1 \over \exp({h\nu \over k T_{*}}) - 1 } \right)^{1/2} \ \ \ \label{eqn:sntrans2}
\end{eqnarray}
where we approximated the flux density from the star as a blackbody in (\ref{eqn:sntrans2}).

Using the same approach as in \S\ref{sec:directsamples} to express the stellar quantities, $T_{*}$ and $r_{*}$, in terms of the stellar mass, setting $x=4$ from (\ref{eqn:earthx}), adopting $\bar{\mu} = 30$\,amu and $\rho = 5$\,g\,cm$^{-3}$, the signal-to-noise ratio for a transit observation is:
\begin{eqnarray}
SN_{tr} & = & 240\, {d_{8} \over D_{\rm pc}} \left({T_{p} \over T_{\oplus}}\right)^{2} m^{-1.5} \nonumber \\
&& \times\ \left(\exp(2.5 m^{-0.5}) - 1\right)^{-1/2}
\label{eqn:sntrans4}
\end{eqnarray}

The signal-to-noise ratio for transit observations of terrestrial planets in the habitable zones around stars increases rapidly with decreasing stellar mass: small stars are strongly favored over massive stars at all distances, both because the transit depths are larger and the observation times are longer, more than offsetting the increased photon rate with increasing stellar temperature. This is a general result that also applies to the photometric detection planetary transits, not just the highly challenging observation of atmospheric lines. Notice also that the signal-to-noise ratio depends linearly on $d_{8} / D$, a slower rate than for background limited photometric observations of (\ref{eqn:SNcalc1}).

The challenge of using transits to detect atmospheric features is illustrated by evaluating (\ref{eqn:photom2}) for the Earth/Sun system at a distance of 10\,pc. A single transit of the Earth across the diameter of the Sun takes 16 hours during which an 8\,m telescope would capture $3 \times 10^{13}$ photons at a spectral resolution of $1/\Gamma = \Delta\nu / \nu = 0.01$ at 1\,$\mu$m. The fractional uncertainty in the difference in light curves at two frequencies would be $\sigma_{N_{\gamma}} / N_{\gamma} = 2.6 \times 10^{-7}$, yielding a signal-to-noise ratio of 3 on for a feature with $\tau(h_{atm}) = 1$ and $x=4$, such as the oxygen lines. Since this signal-to-noise ratio is too small for most features of interest and the required telescope is large, it will probably be impossible to detect signatures of more distant Earths using transits at these wavelengths. However, large telescopes might be able to detect deep water features with $\tau(h_{atm})\sim 10^{4}$ at wavelengths of a few microns with sufficient collecting area, if the water vapor extends throughout the atmosphere, unlike the situation on Earth.

The greater complication with this technique, however, is that the orbital plane of the planet must be aligned with the line of sight for a transit to be observed. The a priori probability of alignment, assuming orientations uniformly distributed within the celestial sphere and requiring only that the planet graze some part of the stellar surface, is:
\begin{eqnarray}
f & = & {r_{*} \over R_{p} }, \label{transfrac1} \\
& = & 2 \left({T_{p} \over T_{*}}\right)^{2},
\label{eqn:transfrac2}
\end{eqnarray}
where we have used (\ref{eqn:orbitvstemp}) to derive (\ref{eqn:transfrac2}). The fraction, $f$, is much less than unity even for cool stars, meaning we would need many candidate stars to ensure that even a few transiting planets would be observed. 

\subsection{Samples for transit observations}
\label{subsec:transitsamples}
The maximum distance for transit observations of a star is derived from (\ref{eqn:sntrans4}) by solving for $D_{t}$. Then, using (\ref{eqn:PDMF}) for the PDMF, the sample size distribution, $dN_{t} / dm$, is just $(4\pi / 3) D_{t}^{3}(m) \xi(m) $:
\begin{eqnarray}
dN_{t} / dm & = & 1000\, d_{8}^{3} \left({T_{p} \over T_{\oplus}}\right)^{6} \nonumber \\
 &&   \times {m^{-\alpha_{2}} \over \left(\exp(2.5 m^{-0.5}) - 1 \right)^{3/2}} \ \ \ \ \ \ \  
\label{eqn:numftnt}
\end{eqnarray}
with $\alpha_{2} = 5.5$ for $m \le 1$, or $\alpha_{2} = 9.5$ for $m > 1$.  The transit technique strongly favors the lower mass stars and hot planets near the inner edge of the habitable zone. Equation (\ref{eqn:numftnt}) has a strong maximum near m = 0.12 and falls off very steeply above m=1.

One other quantity of interest is the expected number of stars in the sample in which the orbital planes will align so that the planets will transit the face of the star. Using (\ref{eqn:transfrac2}) with the formalism above, the expected number is:
\begin{eqnarray}
dN_{et} / dm & = & f(m) dN_{t} / dm \\
  & = & 2\, d_{8}^{3} \left({T_{p} \over T_{\oplus}}\right)^{8} \nonumber \\
  && \times {m^{-\alpha_{3}} \over \left(\exp(2.5 m^{-0.5}) - 1 \right)^{3/2}} \ \ \ \ \ \ \  
\label{eqn:transfracnum}
\end{eqnarray}
with $\alpha_{3} = 6.5$ for $m \le 1$, or $\alpha_{3} = 10.5$ for $m > 1$. The number of {\it observed} transits is so heavily weighted toward low mass stars that the technique favors those near the nuclear burning limit.

Integrating (\ref{eqn:numftnt}) and (\ref{eqn:transfracnum}) from $m=0.1$ to $\infty$ yields the total sample size and expected number of in the sample in which the orbital plane of the planet is aligned for transits:
\begin{eqnarray}
N_{t} & \approx & 500\, d_{8}^{3} \left({T_{p} \over T_{\oplus}}\right)^{6} \label{eqn:tottransits}     \\
N_{et} & \approx & 5\,  d_{8}^{3} \left({T_{p} \over T_{\oplus}}\right)^{8} \label{eqn:expecttransits}
\end{eqnarray}
The final two columns of Table~\ref{tab:transits} contains the total number of candidates and expected number of observed transits from (\ref{eqn:tottransits}), and (\ref{eqn:expecttransits}). 

Notice that the assumptions used to arrive at (\ref{eqn:expecttransits}) are for the best case of maximum transit time and perfect detection efficiency, and there is even a slight increase in the observing time if the number of transits is not truncated to an integer; this result is the best we can do.

\begin{figure}[htb]
\plotone{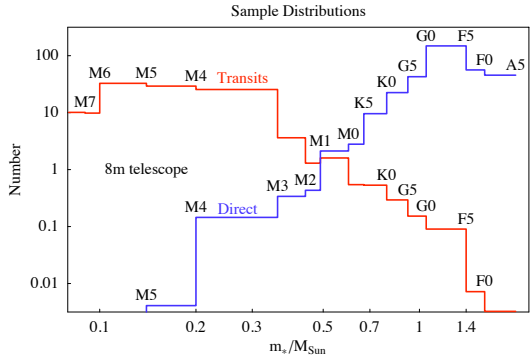}
\caption{The distribution of sample stars vs. spectral type that could be observed with an 8\,m telescope direct detection with coronagraphs from (\ref{eqn:dnddm}) with $f_\theta=2$ (blue) and through transit observations from (\ref{eqn:numftnt}) (red).}
\label{fig:sampledistribution}
\end{figure}

Figure~\ref{fig:sampledistribution} shows how the samples for direct and transit observations are distributed across spectral type. These are complementary techniques, with direct observations favoring stars of about 1\,$M_{\odot}$, and transit observations favoring low mass stars near the nuclear burning limit. It will be useful to use both techniques to characterize the atmospheres of planets around stars of all masses.

These results again point to the great advantage that large telescopes bring to the study of transiting exoplanets. An 8\,m telescope should see of order 5 transiting planets in the terrestrial zones around low-mass stars in the Solar neighborhood, if planets are common, and a 16\,m would yield about 40. The actual number of candidates needed to ensure the detection of a life-bearing planet depends on how commonly they occur, a topic treated in the final section.  

%Figure 4

\section{Samples from Star Catalogs}
\label{sec:catalogs}
There have been several approaches to deriving samples of stars suitable for direct detection of terrestrial planets. One approach is to take all stars within a complete nearby sample, such as those within 8\,pc of the Solar System \cite{rei04} and exclude only the multiple star systems and white dwarfs that would be difficult to observe. Another approach is to apply criteria that maximize the likelihood that some evolved life forms might exist by excluding in addition the very young stars, the evolved giant stars, and stars with a lot of activity. Turnbull (2004) culled 131 such stars from a sample of 2300 within 30\,pc concentrating on ``Sun-like'' F, G, and K stars. A more extensive sample of stars suitable for the SETI program is given by Turnbull \& Tarter (2003) including approximately 18,000 candidates within 2\,kpc. This ``SETI sample'' is important for its inclusion of stars at large distances that could be searched by very large space telescopes, and although they concentrate on solar-type stars, they include an incomplete sample of lower mass, a potentially restrictive limit as shown below.
   
We examined each sample to find stars whose habitable zones can be resolved within the maximum distance for telescopes of different diameters. The observational parameters for these calculations are: $SN=10$, wavelength of 1\,$\mu$m, spectral resolution $\Delta\nu/\nu = 0.01$, detection efficiency $\eta = 1$, observing time $t = 24$\,hr. 

Table~\ref{tab:1lambda} lists the maximum distances and inner working angles derived under these assumptions along with the numbers of stars available for study from the SETI and 8\,pc samples. Table~\ref{tab:3lambda} shows how the numbers change if the inner working angle is assumed to be $3\lambda/d_{tel}$, typical of the best extant coronagraph designs, and allowing for an overall detection efficiency of 0.25. 

\begin{figure}[thb]
\plotone{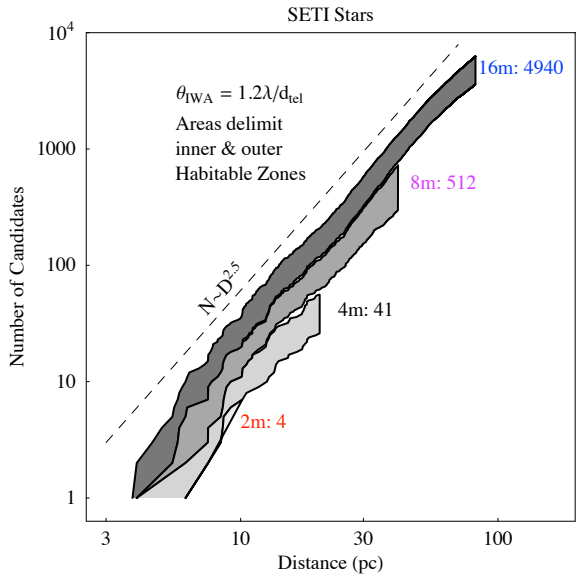}
\caption{Cumulative number of stars from the SETI list vs. $d_{\rm tel}$ using the ideal case of $\eta = 1$ and $f_{\theta} = 1$, as in Table~\ref{tab:1lambda}. Each telescope size shows sample growth in a band between two bounds corresponding to $\theta_{IWA} \le R_{in}$ (lower bound) and $\theta_{IWA} \le R_{out}$ (upper bound). The maximum detection distance from (\ref{eqn:distsense}) forms the right hand boundary of each sample. The numbers next to each diameter are the mean numbers of candidates between the two bands out to the maximum distance, as appear in columns 4 and 5 of the table.}
\label{fig:HC-L}
\end{figure}

Figure~\ref{fig:HC-L} shows the cumulative number of stars whose habitable zones could be studied by single telescopes of different sizes for the SETI sample. The bands of growth delineate the cumulative numbers using either the inner (smaller number) or outer habitable zone radius. The numbers of candidates out to the maximum assumed distances for the different telescope sizes are given in the figure as the average of these two criteria. 

For the SETI sample, the total number of accessible stars increases by an order of magnitude when the telescope diameter doubles. This increase is more rapid than the $d_{\rm tel}^{3}$ rate predicted by the analytical calculations. A close examination of the SETI sample shows that it deviates markedly from the PDMF in its distribution of stars, with a sharp peak in the number at about 0.5\, M$_{\odot}$, a falling distribution toward lower masses and a flattening of the distribution between 1 and 1.5\, M$_{\odot}$,, completely unlike the local PDMF. This distribution over stellar masses also changes with distance, meaning the simple scaling laws derived for a constant PDMF will not describe the behavior of this sample. It is useful to note how these selection effects can distort the sample sizes available for study, when additional criteria are used to select stars from the solar neighborhood that may be suitable for harboring life-bearing planets.

%Figure 5

These figures represent ideal observations, discounting operational inefficiencies, assuming the planet will be at maximum elongation at the time of an observation, ignoring additional background from the presence of interplanetary dust in the distant planetary system, and assuming perfect starlight elimination. If the interplanetary dust in distant planetary systems is typically as strong as it is locally, the effect will be to triple the amount of background light per observation on average in (\ref{eqn:SNcalc2}): $I(\nu, z) \rightarrow 3 I(\nu, z)$, and the increased background effectively decreases $D_{max}$ by a factor of $(1/3)^\frac{1}{4} = 0.8$, decreasing the sample size by about a factor of $(1/3)^\frac{2.5}{4} = 0.5$. If $10^{-9}$ of the starlight leaks into the image of the planet the background increases by nearly an order of magnitude, further limiting the observing distance by a factor 0.56 and decreasing the sample size by a factor of more than four .

The inner working angle is a linear function of observing wavelength. We adopted $\lambda = 1\,\mu$m as the minimum wavelength of interest to pick up spectral features, but it is obvious from Figure~\ref{fig:earthspectrum} that doubling or tripling this wavelength would be an enormous benefit to search for all the molecular signatures of life. Doubling the wavelength would decrease the sample size by another factor of about two for a coronagraph or an interferometer, where the inner working angle depends on the wavelength and the system size. 

Moreover, it is unlikely that a telescope whose inner working angle just resolves the inner habitable zone will be adequate to observe exoplanets around any one star owing to the relatively small fraction of the projected orbit that will subtend large angular displacements for typical orbital inclinations \cite{bro05, bro07}. Realistically, $\theta_{IWA}$ will have to be much smaller than $\theta_{IHZ}$ to ensure a good chance of observing the planet in most cases.

Each of these effects will decrease the sample sizes in the tables by factors of a few, and it is clear that the numbers in the tables might be more than an order of magnitude too large for the practical problem of life signatures. 

On the other hand, the range for terrestrial-mass planets thought to be suitable for life extends to $10\,M_{\oplus}$, meaning the observed area of the planet in (\ref{eqn:fnuplanet}) will be about $10^{2/3} \approx 5$ times larger than $\pi R_{E}^{2}$, increasing the sample size for direct observations by about a factor of 7 ($N \propto D^{2.5} \propto R^{2.5} \propto M^{2.5/3} = 10^{2.5/3} \approx 7$). This factor would boost the numbers in the tables to offset the decreases from other effects above. Furthermore, adopting 24\,hr as the limiting integration time could also be too conservative. For an integration time of 100\,hr, the distance limit would increase by $(\case{100}{24})^{1/4}$, and the sample size by $(\case{100}{24})^{2.5/4} = 2.4$.

We can apply also (\ref{eqn:sntrans4}) to these catalogs to find the number of stars where the atmospheric signatures can be detected through transit experiments. Table~\ref{tab:transits} gives the results. The I magnitudes were used to calculate the photon rates. The 8\,pc sample clearly limits the number of potential low luminosity stars for telescopes larger than a few meters. Using the analysis of \S\ref{subsec:transitsamples}, we can predict the sample of accessible stars (column 6) and the expected number from these samples where the orbits align for transits (column 7). Evidently, the samples for transit observations will be orders of magnitude smaller than the samples for direct observations for all telescopes.

These uncertainties show why refinements to the parameters, for example using the greenhouse effect to increase the size of the habitable zone slightly or trying to estimate the exo-zodiacal light background precisely, can complicate the analysis without bringing any greater insight into the likelihood of detecting life-bearing planets. Moreover, the assumptions used here are optimistic about technological advances for observing exoplanets. There are no margins built in: we assume that all the starlight is rejected, there is no exo-zodiacal background, the detection efficiency is nearly ideal, and there is sufficient resolution when the inner working angle is equal to the size of the habitable zone. The numbers of candidate stars listed in Table~\ref{tab:3lambda} only take into account margins for $\theta_{IWA}$ and detection efficiency, and we believe they may safely be taken as upper limits to the number of observable stars for a given telescope diameter, allowing for some margin in the uncertainties of the properties of exoplanets. Characterizing the atmospheres of Earth-like planets around other stars will be a challenging problem for the foreseeable future.

\section{Likelihood of Life Bearing Planets}
\label{sec:likelihoodlife}
The number of life-bearing planets that can be detected in a survey will depend on the fraction of accessible candidate stars with an Earth-like planet in the habitable zones around the star as well as likelihood that the planets develop life quickly enough so that it starts to dominate the planet's atmospheric chemistry. This fraction is conventionally called $\eta_{\rm Earth}.$\footnote{There are various definitions of $\eta_{\rm Earth}$ in the literature that may not require that the planet is in the habitable zone nor that it has developed life; the definition here is more demanding.}

\begin{figure}[thb]
\plotone{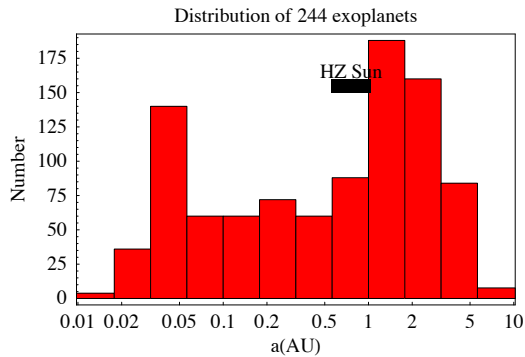}
\caption{The distribution of the semi-major axes of 244 known exoplanets is shown along with the size of the habitable zone around the Sun. Although the location of the habitable zone for each candidate will differ from that of the Sun, the size will be the same in logarithmic units.}
\label{fig:orbitalaxes}
\end{figure}

One estimate of the likelihood that a planet will be found in the habitable zone around a star can be derived from the fraction of stars with known exoplanets and the distribution of their orbital semi-major axes. There are presently 244 known exoplanets\footnote{http://exoplanet.eu/catalog-all.php}, all much more massive than Earth, whose orbital parameters have been estimated. Figure~\ref{fig:orbitalaxes} shows the distribution of semi-major axes of this sample. The range spans approximately 2.5 orders of magnitude. In these logarithmic units, the habitable zone as defined by (\ref{eqn:orbitvslum}) is 0.27 wide, about 10\% of the entire range. Because these exoplanets have been found in only 15\% of the stars surveyed, the empirical chance of finding a massive exoplanet in the habitable zone around a star is 1.5\%. If this analysis also applies to terrestrial-mass planets, a sample of order 100 stars would be required to guarantee ($1\sigma$) at least one terrestrial planet in the habitable zone around a star.  

%Figure 6

This estimate could easily be too low for the problem at hand, however. There are many planets in the Solar System, of which three, Venus, Earth, and Mars, might be capable of supporting life if they are within the habitable zone. The empirical chance of detecting massive exoplanets may be lower than the chance of finding a terrestrial planet, if the Solar System is typical. If planetary systems normally contain several planets of terrestrial mass, 0.5 - 10\,$M_{\oplus}$, the chance that one will be in the habitable zone rises considerably. The microlensing detection of a 5.5\,$M_{\oplus}$ planet around an M dwarf (Beaulieu et al. 2006) suggests that terrestrial planets may be more common around low mass stars than the empirical frequency of more massive planets. An optimist might say that {\it every} planetary system will have a planet within the habitable zone. The question of whether any could support life would depend on their mass; in the solar system, only about 1/3 of the known planets are in the right range. 

How likely is it that life has arisen and evolved to the stage where the atmospheric chemistry reflects its presence? On Earth, life arose quickly and began to alter the atmosphere in ways that might have been detectable from afar by an age of order 1\,Gyr \cite{kalt07}. It took more than 3.5 billion years to produce substantial amounts of oxygen via photosynthetic organisms to a state we would recognize today \cite[and references therein]{kas03}. The oxygen would disappear from the atmosphere in few million years if life were to cease, and CO$_{2}$ would dissolve in the oceans within a few thousand years, eliminating the most prominent atmospheric signature of organisms on Earth. It required a series of unlikely accidents for Earth to alter its atmosphere, suggesting that it may not occur easily on other planets \cite{wb00}.

Thus, an estimate made a priori of the fraction of stars with life-bearing planets depends entirely on how much faith we have that all the circumstances are favorable, that is the number of apparent miracles we are willing to believe. If all stars have planetary systems (but only 15\% have massive planets), and all planetary systems have at least one Earth-like planet within the habitable zone, and this planet always evolves life to dominate its atmospheric chemistry, then nearly 100\% of the suitable stars will have planets indicative of life, i.e. $\eta_{\rm Earth} \sim 1$. Even a 4\,m space telescope might be adequate to carry out the first observations. 

On the other hand, if only 15\% of stars typically have planetary systems of which the likelihood of an Earth-like planet in the terrestrial zone is only 30\% and the chance of evolving life is less than 1, then $\eta_{\rm Earth} < 0.05$, and it may be much less than this value if the sequence of events leading to life-signatures in the atmospheric spectra are more rare than common; many of the low mass stars making up the majority of the complete 8\,pc sample are likely to be too young to have evolved life as on Earth \cite{rei07}. A pessimist might conclude that $\eta_{\rm Earth} << 0.01$. 

We believe that a pragmatic approach to the study of life-bearing exoplanets will require more than 100 candidate stars to yield at least one with the characteristics we seek, requiring a large ($>8$\,m) space telescope. A very large space telescope of order 16\,m diameter would have thousands of candidate stars to study and, while technically challenging to build, would also be an excellent tool for examining exoplanets found by other means, such as spectra-photometry of the transits discussed in \S\ref{sec:transits}.

The range of uncertainty will narrow considerably when the results of the Kepler mission to find Earth-like planets around distant stars are known in a few years. However, Kepler is observing more than $10^{5}$ distant ($>1$\,kpc) stars, very few of which will be close enough to look for atmospheric signatures. In the absence of very large space telescopes to find and study nearby stars, it will still leave open the question about about the likelihood of life outside of the Solar system.

\begin{acknowledgments}
I am grateful to an anonymous referee who suggested several qualitative changes in the approach to this paper, vastly improving it. I also thank Robert Brown, Peter McCullough, James Pringle, Marc Postman, Neill Reid, Kailash Sahu, David Soderblom, and Jeff Valenti for their advice and to Mike Hauser and Matt Mountain for encouragement to pursue this work. This research was supported by NASA through its contract to AURA and the Space Telescope Science Institute and the University of California.
\end{acknowledgments}

%--------------------------------------------------   Tables ------------------------------------------------------------------------
%Table 1
\begin{deluxetable}{cccrrrr}
\tablecaption{Sample sizes with ideal telescopes\tablenotemark{a}\label{tab:1lambda}}
\tablewidth{0pt}
\tablehead{
\colhead{$d_{\rm tel}$} & \colhead{$\theta_{IWA}$} & \colhead{$D_{\rm max}$}  & \multicolumn{2}{c}{n(SETI)} & \multicolumn{2}{c}{n(8pc)} \\
\colhead{m} & \colhead{mas} & \colhead{pc} & \colhead{IHZ\tablenotemark{b}} & \colhead{OHZ\tablenotemark{b}}  & \colhead{IHZ\tablenotemark{b}} & \colhead{OHZ\tablenotemark{b}} 
}
\startdata
2 & 123 & 10 & 2 & 7 & 4\ \  & 8 \ \ \ \\
4 & 62 & 20 & 26 & 56 & 9\ \  & 16 \ \ \ \\
8 & 31 & 41 & 300 & 725 & 19\ \  & 29 \ \ \  \\
16 & 15 & 81 & 3607 & 6272 & 30\ \  & 55 \ \ \  \\
\hline
\enddata
\\
\tablenotetext{a}{$\theta_{IWA} = 1.2{\lambda \over d_{tel}}$ @ $1\,\mu$m, $\eta=1$, $t = 24$\,hr, ${\Delta\nu \over \nu} = 0.01$}
\tablenotetext{b}{$\theta_{IWA} \le R_{p}(IHZ)$ or $ R_{p}(OHZ)$}
\end{deluxetable}

%Table 2
\begin{deluxetable}{cccrrrrr}
\tablecaption{Sample sizes with practical telescopes\tablenotemark{a}\label{tab:3lambda}}
\tablewidth{0pt}
\tablehead{
\colhead{$d_{\rm tel}$} & \colhead{$\theta_{IWA}$} & \colhead{$D_{\rm max}$} 
 & \multicolumn{2}{c}{n(SETI)} & \multicolumn{2}{c}{n(8pc)} & \colhead{$\langle n({\rm tot}) \rangle$\tablenotemark{b}} \\
\colhead{m} & \colhead{mas} & \colhead{pc} & \colhead{IHZ} & \colhead{OHZ}  & \colhead{IHZ} & \colhead{OHZ} & \colhead{OHZ} 
}
\startdata
2 & 309 &   6.8  &     0 &     0 &    1 &   1 &          4 \ \ \\
4 & 155 & 13.5 &     3 &   10 &    3 &   7 &        35 \ \ \\
8 &   77 &  27   &   18 &   78 &    7 & 14 &      280 \ \  \\
16 & 39 &  54   & 241 & 1092 & 14 & 22 & 2240 \ \  \\
\hline
\enddata
\\
\tablenotetext{a}{$\theta_{IWA} = 3{\lambda \over d_{tel}}$ @ $1\,\mu$m, $\eta=0.25$, $t = 24$\,hr, ${\Delta\nu \over \nu} = 0.01$}
\tablenotetext{b}{Calculated from (\ref{eqn:Ncorona3}) }
\end{deluxetable}

%Table 3
\begin{deluxetable}{cccccrr}
\tablecaption{Sample sizes for transit detections\tablenotemark{a}\label{tab:transits}}
\tablewidth{0pt}
\tablehead{
\colhead{$d_{\rm tel}$(m)} & \multicolumn{2}{c}{n(SETI)} & \multicolumn{2}{c}{n(8pc)\tablenotemark{b}} & \colhead{$n_{\rm tot}$\tablenotemark{d}} &  \colhead{$\langle n_{\rm expect} \rangle $\tablenotemark{e}} \\
   & \colhead{$R_{\rm OHZ}$\tablenotemark{b}} & \colhead{$R_{\rm IHZ}$\tablenotemark{c}} & \colhead{$R_{\rm OHZ}$} & \colhead{$R_{\rm IHZ}$}  &  \colhead{$T_{p}=T_{\oplus}$} & \colhead{$T_{p}=T_{\oplus}$}
}
\startdata
2 &     0 &  1 &  3 & 19 &  8 &  0 \ \ \  \\
4 &     3 &  33 &  22 & 71 & 62 & 1 \ \ \ \\
8 &    40 &  256 &   75 & 87 & 500 & 5 \ \ \  \\
16 &  284 & 942 &   87 & 90 &  4000 & 40 \ \ \  \\
\hline
\enddata
\\
\tablenotetext{a}{$\Delta\nu / \nu = 0.01$, $10\sigma$, 5\,yr, $\rho = 5.5\,$g\,cm$^{-3}$, $\bar{\mu} = 30$\,amu, $x=4.2$}
\tablenotetext{b}{Numbers restricted by 8\,pc sample limit}
\tablenotetext{c}{Planet orbit at inner or outer HZ: $T_{p} = 373$ or 273\,K}
\tablenotetext{d}{Calculated from (\ref{eqn:tottransits})}
\tablenotetext{3}{Calculated from (\ref{eqn:expecttransits})}
\end{deluxetable}

\end{document}